\documentclass[opre,nonblindrev]{informs3}

\SingleSpacedXI

\usepackage{endnotes}
\let\footnote=\endnote

\usepackage{amsmath}
\usepackage{bbm}
\usepackage{amsfonts}
\usepackage{algorithm}
\usepackage{algpseudocode}
\usepackage{hyperref}
\usepackage{appendix}
\usepackage{mathrsfs}

\usepackage{natbib}
 \bibpunct[, ]{(}{)}{,}{n}{}{,}%
 \def\bibfont{\small}%

\TheoremsNumberedThrough
\ECRepeatTheorems
\EquationsNumberedThrough

\begin{document}

\RUNAUTHOR{Shi and Mehrotra}

\RUNTITLE{On Approximation of Robust Max-Cut and Related Problems using Randomized Rounding Algorithms}

\TITLE{On Approximation of Robust Max-Cut and Related Problems using Randomized Rounding Algorithms}

\ARTICLEAUTHORS{
\AUTHOR{Haoyan Shi}
\AFF{Department of Mathematics and Department of Computer Science, Northwestern University, Evanston, IL 60208, \EMAIL{haoyanshi2025@u.northwestern.edu}} 
\AUTHOR{Sanjay Mehrotra}
\AFF{Department of Industrial Engineering and Management Sciences, Northwestern University, Evanston, IL 60208, \EMAIL{mehrotra@northwestern.edu}}
}

\ABSTRACT{
Goemans and Williamson proposed a randomized rounding algorithm for the MAX-CUT problem with a 0.878 approximation bound in expectation. The 0.878 approximation bound remains the best-known approximation bound for this APX-hard problem. Their approach was subsequently applied to other related problems such as Max-DiCut, MAX-SAT, and Max-2SAT, etc. We show that the randomized rounding algorithm can also be used to achieve a 0.878 approximation bound for the robust and distributionally robust counterparts of the max-cut problem. We also show that the approximation bounds for the other problems are maintained for their robust and distributionally robust counterparts if the randomization projection framework is used.
}

\KEYWORDS{approximation algorithm; maximum cut; robust optimization} 
\maketitle

\section{Introduction}
Let $G = (V, E)$ be an undirected graph with vertex index set $V$, and edge set $E$. The classical max-cut problem partitions $V$ into two sets, $S \subseteq V$ and $\overline{S} = V \setminus S$, such that the sum the non-negative weights, $w_{ij}=w_{ji}, i, j\in V$ passing through the partition is maximized. The max-cut problem can be written as a binary quadratic program: 
\begin{equation}
\begin{array}{rl}
\text{Val(P)} \stackrel{\text{def}}{=}\max\limits_{y \in \{-1, 1\}^{|V|}} & \frac{1}{2} \sum\limits_{i<j}w_{ij}(1-y_iy_j), 
\end{array}
\label{max-cut}
\tag{P}
\end{equation}
where $y_i = 1$ if vertex $i \in S$ and $y_i = -1$ if vertex $i \in \overline{S}$.  The max-cut problem is APX-hard \cite{Papadimitriou}, i.e., it does not admit a polynomial-time approximation algorithm that can obtain a solution of  \eqref{max-cut} to arbitrary precision.  Goemans and Williamson \cite{Goemans01} developed a seminal polynomial-time algorithm giving a $0.878$ approximation bound. The $0.878$ (to be precise $0.87856$) approximation bound remains the best-known approximation bound in expectation. This is the best approximation ratio that can be obtained for the max-cut problem under Koht's unique games conjecture \cite{kho02, khot, brakensiek}. 

In their randomized rounding algorithm, called G-W approach, Goemans and Williamson \cite{Goemans01} first reformulate \eqref{max-cut} in a lifted space, where $\{-1, 1\}$  is mapped to a unit sphere $S_n\stackrel{\text{def}}{=} \{u\in \mathbb{R}^n \;|\;\|u\|^2 = 1\}$. The resulting relaxed model is shown to admit a semidefinite programming formulation. A randomized projection of a solution of the semidefinite program is used to obtain an approximate solution of  \eqref{max-cut}. Specifically, a random vector $r$ uniformly distributed on $S_n$ is used to project a solution vector $u_i$, and $y_i := sgn(r^Tu_i)$.  This approach was subsequently applied to Max Di-Cut, Max-SAT, and Max-2SAT problems \cite{Goemans01}. For example, a $0.79607$ approximation bound was proved for the Max-DiCut problem by Goemans and Williamson \cite{Goemans01}. The Max-DiCut problem considers a directed graph $G = (V, A)$ and finds a set $S \in V$ to maximize the weights of the directed edges whose tails are in $S$ and heads are in $\overline{S} = V \setminus S$. Goemans and Williamson \cite{Goemans01} model the Max-DiCut problem as the binary quadratic program:
\begin{equation}
\begin{array}{rl}
\max\limits_{y \in \{-1, 1\}^{|V|}} &  \sum\limits_{i,j,k}\left[c_{ijk}(1-y_iy_j - y_iy_k + y_jy_k) + d_{ijk}(1 + y_iy_j + y_iy_k + y_jy_k)\right],
\end{array}
\label{max-ThreeTerms}
\tag{P'}
\end{equation}
where $c_{ijk}, d_{ijk} \geq 0$, which is subsequently used for their $0.79607$ approximation result. The $0.79607$ approximation bound was improved to $0.859$ by Feige and Goemans \cite{feige}. Specifically, Feige and Goeman \cite{feige} use an alternative formulation (See Section~\ref{sec:MaxDiCut}) of Max-DiCut and a non-uniform rounding scheme, where $r$ is skewed in view of the formulation of Max-DiCut used in developing the approximation method. The $0.859$ bound has since been improved to $0.87446$ \cite{brakensiek}.

\subsection{Contributions}
This paper considers the robust and distributionally robust counterparts to \eqref{max-cut}. \begin{equation}
\begin{array}{rl}
\text{Val(RP)} \stackrel{\text{def}}{=} \frac{1}{2} \max\limits_{y \in \{-1, 1\}^{|V|}} \min\limits_{W \in {\cal W}\subseteq {\mathbb R}^{n\times n}_+} & \sum\limits_{i<j}w_{ij}(1-y_iy_j),
\end{array}
\label{robust-max-cut}
\tag{RP}
\end{equation}
where ${\cal W}$ is an uncertainty set in $\mathbb{R}^{n\times n}_+$. We also consider the distributionally robust counterpart of \eqref{max-cut}, which lets the weight matrix to be random on a non-negative support and allows it to follow some distribution $\mathbb{P}_W$ from an ambiguity set ${\cal P}_+$:
\begin{equation}
\begin{array}{rl}
\text{Val(DRP)} \stackrel{\text{def}}{=} \frac{1}{2} \max\limits_{y \in \{-1, 1\}^{|V|}} \min\limits_{\tilde{W}\sim \mathbb{P}_W \in {\cal P}_+} & \mathbb{E}_{\mathbb{P}_W}\left[\sum\limits_{i<j}\tilde{w}_{ij}(1-y_iy_j)\right], 
\end{array}
\label{dro-max-cut}
\tag{DRP}
\end{equation}
where the random weights $\tilde{W}=[\tilde{w}_{ij}]$ follow an unknown probability distribution $\mathbb{P}_W$ from an ambiguous set of random matrices supported on $\Xi \subseteq \mathbb{R}_+^{V\times V}$.

We show that the G-W algorithm for \eqref{robust-max-cut} and \eqref{dro-max-cut} achieves the same $0.878$ approximation ratio as for \eqref{max-cut}. We further use robust and distributionally robust counterparts of \eqref{max-ThreeTerms} to emphasize that the applicability of the G-W randomized approximation approach extends to more general settings without compromising the approximation ratio achieved in the deterministic model. While we prove the results under the settings described in Goemans and Williamson~\cite{Goemans01}, we note that the results are not dependent on how the random vector $r$ is generated. We also note that the results are not limited to the max-cut problem. They are also applicable to robust and distributionally robust counterparts of the related problems that use SDP relaxation and randomized rounding schemes, including maximum directed cut (Max Di-Cut) and maximum k-constraint satisfaction problem (Max k-CSP) (also see Appendix~A for special cases of max-cut from \cite{Goemans01}).  We discuss this in a general framework. 

Throughout the paper we assume that the robust optimization models are well-posed with finite optimal value, and a solution to the inner problem in the robust models exists. We also assume the expected values arising in the analysis are well-defined.

\subsection{Organization}
This paper is organized as follows. We first review the G-W algorithm for \eqref{max-cut} in Section~\ref{sec:GoemanWilliamson}. We provide some preliminary results for robust and distributionally robust generalizations of the max cut problem in Section~\ref{sec:BasicResultsRobustMaxCut}, where we also state the G-W algorithm for \eqref{robust-max-cut} and \eqref{dro-max-cut}. We analyze the approximation ratio for robust and distributionally robust counterparts of the G-W algorithm in Section~\ref{sec:Analysis}. We extend our results to Max Di-Cut and present a generalized framework of analysis in robust settings in Section~\ref{sec:Generalizations}. Lastly, we discuss the tractability of the G-W algorithm for the robust and distributionally robust counterparts in Section~\ref{sec:Tractability}. We further discuss the generalizability of the approximation bound to other combinatorial optimization problems in Section~\ref{sec:ConcludingRemarks}.

\section{Goemans-Williamson Algorithm for Robust Counterparts}
\label{sec:GoemanWilliamson}
Goemans and Williamson consider the following relaxation of \eqref{max-cut}:
\begin{equation}
\begin{array}{rl}
\max\limits_{u_i \in S_n, \forall i \in V} &  \frac{1}{2} \sum\limits_{i<j}w_{ij}(1-u_i \cdot u_j),  
\end{array}
\label{relax-max-cut}
\tag{P-R}
\end{equation}
which has the following SDP equivalent reformulation:
\begin{equation}
\begin{array}{rl}
\frac{1}{4} \max\limits_{Y \succeq 0, Y_{ii} = 1, \forall i \in V} & \langle W, {\bf 1} - Y \rangle , 
\end{array}
\label{sd-max-cut}
\tag{P-SD}
\end{equation}
where $Y \in \mathbb{R}^{n \times n}$ and ${\bf 1} \in \mathbb{R}^{n \times n}$ denotes a matrix with $1's$ for all entries. From an optimal solution $\hat{Y}$ of  \eqref{sd-max-cut}, we first performs Cholesky factorization to obtain $u_i, i \in V$ for \eqref{relax-max-cut} by letting $u_i = Ye_i$, where $e_i$ is the $i$th column of an identity matrix. G-W algorithm now draws a random vector $r$ using a uniform distribution on $S_n$. It sets $y_i = 1$ if $u_i \cdot r \geq 0$, and $y_i = -1$ otherwise. This rounding scheme is equivalent to drawing a sample from a Gaussian distribution ${\cal N}(0,\hat{Y})$ \cite{BertsimasYe}. The G-W approximation algorithm is summarized in Algorithm~1. 

\begin{algorithm} \label{alg:GW-Algorithm}
\caption{Goemans and Williamson Approximation Algorithm} 
\begin{algorithmic}[1]
\State \textbf{Step 1:} Solve \eqref{sd-max-cut} and obtain an optimal solution $\hat{Y}$ 
\State \textbf{Step 2:} Perform Cholesky factorization of $\hat{Y}$ to obtain {$\hat{u}_i, i \in V$. 
\State \textbf{Step 3:} Generate a sample $r \in S_n$ following uniform distribution $\mathbb{U}$ on $S_n$.
\State \textbf{Step 4:} Set $\hat{y}_i = 1$ if $\hat{u}_i \cdot r \geq 0$, otherwise set $\hat{y}_i = -1$ for each $i \in V$.
}
\end{algorithmic}
\end{algorithm}

\subsection{Goemans-Williamson Algorithm for Robust Counterparts}
\label{sec:BasicResultsRobustMaxCut}
\subsubsection{Robust Max-Cut.} 
Let us consider the following relaxation of \eqref{robust-max-cut}:
\begin{equation}
\begin{array}{rl}
\text{Val(RP-R)} \stackrel{\text{def}}{=} \frac{1}{2} \max\limits_{u_i \in S_n, \forall i \in V} \min\limits_{W \in {\cal W} \subseteq {\mathbb R}^{n\times n}_+} & \sum\limits_{i<j}w_{ij}(1-u_i \cdot u_j). 
\end{array}
\label{v-robust-max-cut}
\tag{RP-R}
\end{equation}
The SDP reformulation of \eqref{v-robust-max-cut} is given as:
\begin{equation}
\begin{array}{rl}
\text{Val(RP-SD)} \stackrel{\text{def}}{=} \frac{1}{4} \max\limits_{Y \succeq 0, Y_{ii} = 1, \forall i \in V} \min\limits_{W \in {\cal W}\subseteq {\mathbb R}^{n\times n}_+} &\langle W, {\bf 1} - Y \rangle.
\end{array}
\label{SDP-MAXCUT}
\tag{RP-SD}
\end{equation}

\begin{proposition} 
The model \eqref{v-robust-max-cut} is an outer relaxation of \eqref{robust-max-cut}, and \eqref{v-robust-max-cut} is equivalent to \eqref{SDP-MAXCUT}. Also, \text{Val(RP-R)} = \text{Val(RP-SD)} $\geq$ \text{Val(RP)}. 
\label{theorem3}
\end{proposition}
\noindent {\em Proof}. Let $y \in \mathbb{R}^n$ be a feasible solution of \eqref{robust-max-cut}. We can take $u_i = (y_i, 0, \dots, 0) \in \mathbb{R}^n$ for each $i \in V$ and observe that such $u_i$'s are feasible in \eqref{v-robust-max-cut}, and $u_i \cdot u_j = y_iy_j$. The inner minimization problem with set ${\cal W}$ is unchanged. Thus, \eqref{v-robust-max-cut} is an outer relaxation of \eqref{robust-max-cut}. Moreover, $\text{Val(RP-R)} \geq \text{Val(RP)}$. Now in \eqref{SDP-MAXCUT} since $Y \succeq 0$, there exists a matrix $U \in \mathbb{R}^{n \times n}$ such that $Y = U^TU$. This matrix can be obtained by the Cholesky factorization of $Y$. For a fixed $W \in {\cal W}$, the columns of $U$ correspond to the feasible solutions of \eqref{v-robust-max-cut}. Similarly, we can use a feasible solution of \eqref{v-robust-max-cut}, form $U = [u_1,\ldots, u_{|V|}]$, and use $Y=U^TU$ to obtain a feasible solution of \eqref{SDP-MAXCUT}. Since the objective function of the inner minimization problems in \eqref{v-robust-max-cut} and \eqref{SDP-MAXCUT} evaluate to the same value and the feasible set of $W$ has not changed, \text{Val(RP-R)} = \text{Val(RP-SD)}.
\qed

The G-W algorithm for the robust max-cut problem is given in Algorithm~\ref{alg:GW-RP}. 
\begin{algorithm}
\caption{Approximation Algorithm for Robust Max Cut}
\label{alg:GW-RP}
\begin{algorithmic}[1]
\State {\bf Step 1:} Solve \eqref{SDP-MAXCUT}, obtaining optimal $\hat{Y}$ and worst-case weight matrix $\hat{W}$.
\State {\bf Step 2:} Perform Steps 2-4 of Algorithm~1.
\end{algorithmic}
\end{algorithm}
In the next section we will show that this algorithm has $0.878$ approximation ratio in expectation for \eqref{robust-max-cut}.

\subsubsection{Distributionally Robust Max-Cut.}
Analogous to the robust case, we have a relaxation of \eqref{dro-max-cut} as:
\begin{equation}
\begin{array}{rl}
\text{Val(DRP-R)} \stackrel{\text{def}}{=} \frac{1}{2} \max\limits_{u_i \in S_n, \forall i \in V} \min\limits_{\mathbb{P}_W \in {\cal P}_+} & \mathbb{E}_{\mathbb{P}_W}\left[\sum\limits_{i<j} \Tilde{w}_{ij}(1-u_i \cdot u_j)\right], 
\end{array}
\label{v-dro-max-cut}
\tag{DRP-R}
\end{equation}
with the reformulation: 
\begin{equation}
\begin{array}{rl}
\text{Val(DRP-SD)} \stackrel{\text{def}}{=} \frac{1}{4} \max\limits_{Y \succeq 0, Y_{ii} = 1, \forall i \in V} \min\limits_{\mathbb{P}_W \in {\cal P}_+} & \mathbb{E}_{\mathbb{P}_W}\left[\langle \Tilde{W}, \mathbf{1} - Y \rangle \right]. 
\end{array}
\label{SDP-dro-MAXCUT}
\tag{DRP-SD}
\end{equation}

\begin{proposition}
The model \eqref{v-dro-max-cut} is an outer relaxation of \eqref{dro-max-cut}. Moreover, \eqref{v-dro-max-cut} is equivalent to \eqref{SDP-dro-MAXCUT}. Also, \text{Val(DRP-R)} = \text{Val(DRP-SD)} $\geq$ \text{Val(DRP)}.   
\label{prop5}
\end{proposition}
\noindent {\em Proof}. Similar to the proof of Proposition \ref{theorem3}. 
\qed

The approximation algorithm for distributionally robust Max-Cut is similar to Algorithm \ref{alg:GW-RP}, with the first line replaced by the solution of \eqref{SDP-dro-MAXCUT}.
\begin{algorithm}
\caption{Approximation Algorithm for Distributionally Robust Max Cut}
\label{alg:GW-DRP}
\begin{algorithmic}[1]
\State {\bf Step 1:} Solve \eqref{SDP-dro-MAXCUT}, obtaining optimal $\hat{Y}$ and worst weight probability distribution $\mathbb{P}_{\hat{W}}$.
\State {\bf Step 2:} Perform Steps 2-4 of Algorithm~1.
\end{algorithmic}
\end{algorithm}

\section{Analysis of the G-W Algorithm for Robust Counterparts}
\label{sec:Analysis}
The analysis in Goemans and Williamson for the max-cut problem uses the following lemmas, which also play a central role in our proofs.

\begin{lemma} \cite[Lemma 3.2]{Goemans01} \label{lem:GWProbLemma} Let $r$ be a vector uniformly distributed on the unit sphere $S_n$, and $u_i, u_j$ be any vectors belonging to $S_n$. Then,
$\mathbb{P}\left[sgn(u_i \cdot r) \neq sgn(u_j \cdot r)\right] = \frac{1}{\pi} arccos(u_i\cdot u_j)$. 
\end{lemma}
\begin{lemma} \cite[Lemma 3.4, Lemma 3.5]{Goemans01} \label{lem:GWCosecLemma} For $-1 \leq y \leq 1$, $arccos(y)/\pi \geq \frac{\alpha}{2}(1-y),$ where $\alpha > 0.878$. 
\end{lemma}

\subsection{Approximation Bound for Robust Maximum Cut}
Let $(\hat{y},\hat{W})$ be a solution  obtained by Algorithm~\ref{alg:GW-RP}, and  for any $W \in {\cal W}$, $\hat{C}(W) = \sum\limits_{i<j}w_{ij} \frac{(1-\hat{y}_i\hat{y}_j)}{2}$. Let $\mathbb{E}_{\mathbb U}[\hat{C}(W)]$ denote the expected value of $\hat{C}(W)$. The following theorem shows that the expected value of the cut given by $\hat{y}$ is a $0.878$ approximation of the optimal value of \eqref{robust-max-cut}. 

\begin{theorem} \label{thm:RobustMaxCutApproximationTheorem} $\mathbb{E}[\hat{C}(W)] \geq 0.878 \cdot \text{Val\eqref{robust-max-cut}}, \forall\, W \in {\cal W}$. Specifically, $\mathbb{E}[\hat{C}(\hat{W})] \geq 0.878 \cdot \text{Val\eqref{robust-max-cut}}$. 
Moreover, $\text{Val\eqref{robust-max-cut}} \geq \mathbb{E}[\hat{C}(\hat{W}')]$, where $\hat{W}' \stackrel{\text{def}}{=} \arg\min\limits_{W \in {\cal W}} \sum\limits_{i<j} w_{ij} \frac{(1-\hat{y}_i\hat{y}_j)}{2}$. 
\label{theorem1}
\end{theorem}

\noindent {\em Proof}. The expected value of the cut generated by G-W algorithm and evaluated for a $W \in {\cal W}$ is:
\begin{align*}
\mathbb{E}_{\mathbb U}[\hat{C}(W)] & = \mathbb{E}_{\mathbb U}\left[\sum\limits_{i<j}w_{ij} \frac{(1-\hat{y}_i\hat{y}_j)}{2}\right] = \sum\limits_{i<j}w_{ij} \mathbb{E}_{\mathbb{U}}\left[\frac{(1-\hat{y}_i\hat{y}_j)}{2}\right] = \sum\limits_{i<j}w_{ij} \mathbb{P}\left[sgn(\hat{u}_i \cdot r) \neq sgn(\hat{u}_j \cdot r)\right]\\
     & = \sum\limits_{i < j} w_{ij} \frac{\arccos(\hat{u}_i \cdot \hat{u}_j)}{\pi} \qquad \qquad \qquad \qquad \qquad \qquad ({\text by}\ Lemma~\ref{lem:GWProbLemma}) \\
     & \geq 0.878 \sum\limits_{i < j} w_{ij} \left(\frac{1-\hat{u}_i \cdot \hat{u}_j}{2}\right) \qquad \qquad \qquad \qquad ({\text by}\ Lemma~\ref{lem:GWCosecLemma}, \text{ and } w_{ij} \geq 0) \\
     & \geq 0.878 \sum\limits_{i < j} \hat{w}_{ij} \left(\frac{1-\hat{u}_i \cdot \hat{u}_j}{2}\right) \tag{since $\hat{W}$ is the worst-case weights for $\hat{u}_i, i \in V$}\\
     & = 0.878 \, \text{Val\eqref{v-robust-max-cut}} \geq 0.878 \, \text{Val\eqref{robust-max-cut}} \tag{by Proposition \ref{theorem3}}.
\end{align*}
Let $y^*_i, i \in V$ and $W^*$ denote the optimal solution to \eqref{robust-max-cut}. Now, 
\begin{align*}
\mathbb{E}_{\mathbb U}[\hat{C}(\hat{W}')] & = \mathbb{E}_{\mathbb U}\left[\sum\limits_{i<j}\hat{w}'_{ij}\frac{(1-\hat{y}_i\hat{y}_j)}{2}\right] 
 \leq \mathbb{E}_{\mathbb U}\left[\sum\limits_{i<j}w^*_{ij}\frac{(1-y^*_i y^*_j)}{2}\right]  \tag{because of optimality} \\
& = \mathbb{E}_{\mathbb U}[\text{Val\eqref{robust-max-cut}}]  = \text{Val\eqref{robust-max-cut}}. \qed
\end{align*}

Theorem \ref{theorem1} shows that the approximation bound of Goemans and Williamson's algorithm can be generalized to the robust counterpart of the max-cut problem. For all $W \in {\cal W}$, the expected worst-case value of our generated cut evaluated on $W$ is at least 0.878 times the worst-case value of the robust optimal cut, providing a guarantee on the robustness of the generated cut. 

\subsection{Approximation Bound for Distributionally Robust Maximum Cut}
We now analyze the G-W algorithm for \eqref{dro-max-cut}. Let $(\hat{y},\mathbb{P}_{\hat{W}})$ be a solution  obtained by Algorithm~\ref{alg:GW-RP}. For a sample $W \in \Xi$, let $\hat{C}(W) = \sum\limits_{i<j}w_{ij} \frac{(1-\hat{y}_i\hat{y}_j)}{2}$, and $\mathbb{E}_{\mathbb U}[\hat{C}(W)]$ denote the expected value of $\hat{C}(W)$. Now for $\mathbb{P}_W \in {\cal P}_+$, the objective value of \eqref{dro-max-cut} is given by $\mathbb{E}_{\mathbb{P}_W}\left[\mathbb{E}_{\mathbb U}[\hat{C}(W)]\right]$. The proof for the $0.878$ approximation result in the following theorem follows steps similar to those in the proof of Theorem~\ref{thm:RobustMaxCutApproximationTheorem}, while making the observation that the $\tilde{W}\sim {\mathbb{P}_W}$ is independent of $r\sim {\mathbb U}$. 

\begin{theorem}
$\mathbb{E}_{\mathbb{P}_W}\left[\mathbb{E}_{\mathbb U}[\hat{C}(W)]\right] \geq 0.878 \cdot \text{Val(DRP)}, \forall\, \mathbb{P}_W \in {\cal P_+}$. Specifically,  $\mathbb{E}_{\mathbb{P}_{\hat{W}}}\left[\mathbb{E}_{\mathbb U}[\hat{C}(W)]\right] \geq 0.878 \cdot \text{Val(DRP)}$ . Moreover, $\text{Val(DRP)} \geq \mathbb{E}_{\hat{\mathbb{P}}'_W}\left[\mathbb{E}_{\mathbb U}[\hat{C}(W)]\right]$, where $\hat{\mathbb{P}}'_W \stackrel{\text{def}}{=} \arg\min\limits_{\tilde{W}\sim \mathbb{P}_W \in {\cal P_+}} \mathbb{E}_{\mathbb{P}_W} \left[\sum\limits_{i<j} \tilde{w}_{ij} \frac{(1-\hat{y}_i\hat{y}_j)}{2}\right]$. 
\label{theorem7}
\end{theorem}
\noindent {\em Proof}. Let $\mathbb{P}_W \in {\cal P}_+$, and the expected value from Algorithm~\ref{alg:GW-DRP} evaluated using $\tilde{W} \sim \mathbb{P}_W$ is
\begin{align*}
\mathbb{E}_{\mathbb{P}_W}\left[\mathbb{E}_{\mathbb U}[\hat{C}(W)]\right]
&= \sum\limits_{i<j} \mathbb{E}_{\mathbb{P}_W}[w_{ij}] \cdot \mathbb{E}_{\mathbb U}\left[\frac{(1-\hat{y}_i\hat{y}_j)}{2}\right] \tag{by independence} \\ &= \sum\limits_{i<j} \mathbb{E}_{\mathbb{P}_W}[w_{ij}] \cdot \mathbb{P}\left[sgn(\hat{u}_i \cdot r) \neq sgn(\hat{u}_j \cdot r)\right] 
     = \sum\limits_{i < j} \mathbb{E}_{{\mathbb{P}_W}}[w_{ij}] \cdot \frac{\arccos(\hat{u}_i \cdot \hat{u}_j)}{\pi} \tag{by Lemma \ref{lem:GWProbLemma}}\\
     & \geq 0.878 \sum\limits_{i < j} \mathbb{E}_{{\mathbb{P}_W}}[w_{ij}] \cdot \left(\frac{1-\hat{u}_i \cdot \hat{u}_j}{2}\right) \tag{by Lemma \ref{lem:GWCosecLemma}, and $w_{ij} \geq 0$} \\
     & = 0.878 \cdot \mathbb{E}_{{\mathbb{P}_W}}\left[\sum\limits_{i < j} w_{ij} \left(\frac{1-\hat{u}_i \cdot \hat{u}_j}{2}\right)\right] 
     \geq 0.878 \cdot \mathbb{E}_{{\mathbb{P}_{\hat{W}}}}\left[\sum\limits_{i < j} \hat{w}_{ij} \left(\frac{1-\hat{u}_i \cdot \hat{u}_j}{2}\right)\right] \tag{since $\mathbb{P}_{\hat{W}}$ is  worst for $\hat{u}_i, i \in V$}\\
     & = 0.878 \, \text{Val(DRP-R)} \geq 0.878 \, \text{Val(DRP)} \tag{by proposition \ref{prop5}}.
\end{align*}
Let $y^*_i, i \in V$ and $\mathbb{P}_W^*$ denote the optimal solution to \eqref{dro-max-cut}. Thus, 
\begin{align*}
\mathbb{E}_{\hat{\mathbb{P}}'_W}\left[\mathbb{E}_{\mathbb U}[\hat{C}(W)]\right] & = \mathbb{E}_{\hat{\mathbb{P}}_W'} \left[ \mathbb{E}_{\mathbb U}\left[\sum\limits_{i<j}\hat{w}'_{ij}\frac{(1-\hat{y}_i\hat{y}_j)}{2}\right]\right]
 \leq \mathbb{E}_{\mathbb{P}_W^*} \left[\mathbb{E}_{\mathbb U}\left[\sum\limits_{i<j}w^*_{ij}\frac{(1-y^*_i y^*_j)}{2}\right]\right]  \tag{because of optimality} \\
& = \mathbb{E}[\text{Val(DRP)}] = \text{Val(DRP)}. \qed
\end{align*} 

\subsection{The Robust Max-DiCut Problem}
\subsubsection{Basic Formulation.} 
Similar to the \eqref{robust-max-cut}, \eqref{max-ThreeTerms} has a robust counterpart
\begin{equation}
\begin{array}{rl}
\text{Val(RP')} \stackrel{\text{def}}{=} \max\limits_{y \in \{-1, 1\}^{|V|}} \min\limits_{C, D \in {\cal W}\subseteq {\mathbb R}^{n\times n\times n}_+} & \sum\limits_{i,j,k}\left[c_{ijk}(1-y_iy_j - y_iy_k + y_jy_k) + d_{ijk}(1 + y_iy_j + y_iy_k + y_jy_k)\right],
\end{array}
\label{dicut-robust}
\tag{RP'}
\end{equation}
which we can relax to
\begin{equation}
\begin{array}{rl}
\text{Val(RP'-R)} \stackrel{\text{def}}{=} \max\limits_{u_i \in S_n, \forall i \in V} \min\limits_{C, D \in {\cal W}\subseteq {\mathbb R}^{n\times n\times n}_+} & \sum\limits_{i,j,k}\left[c_{ijk}(1 - u_i \cdot u_j - u_i \cdot u_k + u_j \cdot u_k) + d_{ijk}(1 + u_i \cdot u_j + u_i \cdot u_k + u_j \cdot u_k)\right]
\end{array}
\label{max-ThreeTerms-relax}
\tag{RP'-R}
\end{equation}
 Similar to propositions \ref{theorem3} and \ref{prop5}, \eqref{max-ThreeTerms-relax} is an outer relaxation of \eqref{dicut-robust} and Val(RP'-R) $\geq$ Val(RP').

\begin{lemma} \cite[Lemma 7.3.1]{Goemans01} \label{lem:GW-Di=ProbLemma} Let $r$ be a vector uniformly distributed on the unit sphere $S_n$, and $u_i, u_j, u_k$ be any vectors belonging to $S_n$. Then,
$\mathbb{P}\left[sgn(u_i \cdot r) = sgn(u_j \cdot r) = sgn(u_k \cdot r)\right] = 1 - \frac{1}{2\pi} (arccos(u_i\cdot u_j) + arccos(u_i\cdot u_k) + arccos(u_j\cdot u_k))$.
\end{lemma}
\begin{lemma} \cite[Lemma 7.3.2]{Goemans01} \label{lem:GW-Di-CosecLemma} $\forall u_i, u_j, u_k \in S_n$, $1 - \frac{1}{2\pi} (arccos(u_i\cdot u_j) + arccos(u_i\cdot u_k) + arccos(u_j\cdot u_k)) \geq \frac{\beta}{4}(1 + u_i \cdot u_j + u_i \cdot u_k + u_j \cdot u_k)$, where $\beta > 0.796$.
\end{lemma}
 
The $0.79607$ approximation ratio for \eqref{max-ThreeTerms} also holds in its robust counterpart \eqref{dicut-robust} because the proof of approximation bound hinges on Lemmas \ref{lem:GW-Di=ProbLemma} and \ref{lem:GW-Di-CosecLemma}, which do not depend on the uncertain weights. A similar result holds for the distributional robust counterpart of this model. 

\subsubsection{Special Formulation with Biased Rounding.} 
\label{sec:MaxDiCut}
The G-W $0.79607$ approximation bound for Max-DiCut was improved by Feige and Goemans, Matuura and Matsui, and Lewin, Livnat and Zwick \cite{feige, Shiro, Lewin}. The improvement in approximation bounds is due to the biased rounding techniques instead of using the uniform randomized rounding. Numerical experiments were used to show the implicit relationship between the probability and the objective function by the improved approximation ratio found. Since the new approximation ratio is independent of our weight matrix, the results from our previous theorems also hold for these improved rounding techniques. To illustrate why this is true, let us consider the following formulation Max Di-Cut problem from Feige and Goemans \cite{feige}: 
\begin{equation}
\begin{array}{rl}
\frac{1}{4} \max\limits_{u_i \in \mathbb{R}^{|V|+1}, \| u_i \|^2 = 1, i = 0, \ldots, n} & \sum\limits_{i, j} w_{ij} (1 + u_0 \cdot u_i - u_0 \cdot u_j - u_i \cdot u_j) \\
\textit{s.t.} & u_0 \cdot u_i + u_0 \cdot u_j + u_i \cdot u_j \geq -1, \hspace{0.3cm} i,j = 1, \ldots, n \\
& - u_0 \cdot u_i - u_0 \cdot u_j + u_i \cdot u_j \geq -1, \hspace{0.3cm} i,j = 1, \ldots, n \\
& - u_0 \cdot u_i + u_0 \cdot u_j - u_i \cdot u_j \geq -1, \hspace{0.3cm} i,j = 1, \ldots, n \\
& u_0 \cdot u_i - u_0 \cdot u_j - u_i \cdot u_j \geq -1, \hspace{0.3cm} i,j = 1, \ldots, n. 
\end{array}
\label{feige-simple}
\tag{F}
\end{equation}
Let $\mathbb{P}(u_i, u_j)$ be the probability that the edge $(i , j)$ is in the cut after some rounding procedure (may be biased). Let ${\cal U}$ be the set of all $(u_i, u_j)$ pairs that are feasible in \eqref{feige-simple}, and we find the approximation ratio $\alpha \stackrel{\text{def}}{=} \min\limits_{(u_i, u_j) \in {\cal U}} \frac{\mathbb{P}(u_i, u_j)}{\frac{1}{4}(1 + u_0 \cdot u_i - u_0 \cdot u_j - u_i \cdot u_j)}$ numerically \cite{Lewin}. Thus, $\mathbb{P}(u_i, u_j) \geq \frac{\alpha}{4} (1 + u_0 \cdot u_i - u_0 \cdot u_j - u_i \cdot u_j)$, which is independent of the weights.  As a consequence, our analysis of their robust counterparts is unaffected by the biased rounding procedure. Extensive experiments to obtain $\alpha$ are not rigorous enough, but they are commonly used due to simplicity. It is possible to construct a rigorous proof using $\mathscr{R}\text{eal}\mathscr{S}\text{earch}$ but this requires more work \cite{computer}. 

\section{Generalizations}
\label{sec:Generalizations}
In general, the approximation ratio from the relaxation and rounding scheme holds for robust counterparts of other problems. Let us first discuss the relaxation and rounding procedure in a more general setting, and consider the problem
\begin{equation}
\begin{array}{rl}
\text{Val(O)} \stackrel{\text{def}}{=} \max\limits_{x \in {\cal X}} & f(x), 
\end{array}
\label{original}
\tag{O}
\end{equation}
which one may relax to 
\begin{equation}
\begin{array}{rl}
\text{Val(R)} \stackrel{\text{def}}{=} \max\limits_{\Tilde{x} \in \Tilde{{\cal X}}} & \Tilde{f}(\Tilde{x}). 
\end{array}
\label{relaxed}
\tag{R}
\end{equation}
After solving \eqref{relaxed} (which will typically be more tractable to solve), one rounds the solution $\Tilde{x}^* \in  \Tilde{{\cal X}}$ back to $\hat{x} \in {\cal X}$ using a randomized scheme. The expected value of the rounded solution is $\mathbb{E}[f(\hat{x})] = g(\Tilde{x}^*) \geq \alpha \Tilde{f}(\Tilde{x}^*) = \alpha \text{Val(R)} \geq \alpha \text{Val(O)}$, where $\alpha$ is the approximation ratio. The first step is constructed by designing a random rounding procedure, and the second step is obtained by an algebraic relationship that transforms some related function $g$ to the form of the objective function of \eqref{relaxed}. When the algorithm is constructive, we automatically get $\text{Val(O)} \geq \mathbb{E}[f(\hat{x})]$ and thus together we have $\text{Val(O)} \geq \mathbb{E}[f(\hat{x})] \geq \alpha \text{Val(O)}$. 
In the robust setting, \eqref{original} becomes
\begin{equation}
\begin{array}{rl}
\text{Val(RO)} \stackrel{\text{def}}{=} \max\limits_{x \in {\cal X}} \min\limits_{\xi \in \Xi} & f(x, \xi), 
\end{array}
\label{original-robust}
\tag{RO}
\end{equation}
which one may relax to 
\begin{equation}
\begin{array}{rl}
\text{Val(RR)} \stackrel{\text{def}}{=} \max\limits_{\Tilde{x} \in \Tilde{{\cal X}}} \min\limits_{\xi \in \Xi} & \Tilde{f}(\Tilde{x}, \xi). 
\end{array}
\label{relaxed-robust}
\tag{RR}
\end{equation}
After solving \eqref{relaxed-robust}, one obtains the optimal solution $\Tilde{x}^*$ and $\Tilde{\xi}^*$ and rounds $\Tilde{x}^* \in \Tilde{{\cal X}}$ back to $\hat{x} \in {\cal X}$ using randomness. The expected value of the rounded solution evaluated on some fixed $\xi \in \Xi$ is $\mathbb{E}[f(\hat{x}, \xi)] = g(\Tilde{x}^*, \xi) \geq \alpha \Tilde{f}(\Tilde{x}^*, \xi) \geq \alpha \Tilde{f}(\Tilde{x}^*, \Tilde{\xi}^*) = \alpha \text{Val(RR)} \geq \alpha \text{Val(RO)}$, where the third step is by definition of a robust optimal solution. When the algorithm is constructive, we have $\text{Val(RO)} \geq \mathbb{E}[f(\hat{x}, \hat{\xi}^*)]$, where $\hat{\xi}^* \stackrel{\text{def}}{=} \arg\min\limits_{\xi} f(\hat{x}, \xi)$. 

In the distributionally robust setting, \eqref{original} becomes
\begin{equation}
\begin{array}{rl}
\text{Val(DRO)} \stackrel{\text{def}}{=} \max\limits_{x \in {\cal X}} \min\limits_{\mathbb{P}_\xi \in {\cal P}} & \mathbb{E}_{\xi \sim \mathbb{P}_\xi} [f(x, \xi)], 
\end{array}
\label{original-dro}
\tag{DRO}
\end{equation}
where one may relax to 
\begin{equation}
\begin{array}{rl}
\text{Val(RDRO)} \stackrel{\text{def}}{=} \max\limits_{\Tilde{x} \in \Tilde{{\cal X}}} \min\limits_{\mathbb{P}_\xi \in {\cal P}} & \mathbb{E}_{\xi \sim \mathbb{P}_\xi} [\Tilde{f}(\Tilde{x}, \xi)]. 
\end{array}
\label{relaxed-dro}
\tag{RDRO}
\end{equation}
After solving \eqref{relaxed-dro}, one obtains the optimal solution $\Tilde{x}^*$ and $\Tilde{\mathbb{P}}_\xi^*$ and rounds $\Tilde{x}^* \in \Tilde{{\cal X}}$ back to $\hat{x} \in {\cal X}$ using randomness. The expected value of the rounded solution evaluated with $\xi$ following some fixed $\mathbb{P}_\xi \in {\cal P}$ is $\mathbb{E}_{\xi \sim \mathbb{P}_\xi}[f(\hat{x}, \xi)] = \mathbb{E}_{\xi \sim \mathbb{P}_\xi}[g(\Tilde{x}^*, \xi)] \geq \alpha \mathbb{E}_{\xi \sim \mathbb{P}_\xi}[\Tilde{f}(\Tilde{x}^*, \xi)] \geq \alpha \mathbb{E}_{\xi \sim \Tilde{\mathbb{P}}_\xi^*}[\Tilde{f}(\Tilde{x}^*, \xi)] = \alpha \text{Val(RDRO)} \geq \alpha \text{Val(DRO)}$ and we have $\text{Val(DRO)} \geq \mathbb{E}_{\xi \sim \hat{\mathbb{P}}_\xi^*}[f(\hat{x}, \xi)]$, where $\hat{\mathbb{P}}_\xi^* \stackrel{\text{def}}{=} \arg\min\limits_{\mathbb{P}_\xi \in {\cal P}} \mathbb{E}_{\xi \sim \mathbb{P}_\xi} [f(\hat{x}, \xi)]$. 

\subsection{Example: Maximum k-Constraint Satisfaction Problem (MAX k-CSP)}
To show that our generalized framework indeed works for robust counterparts of a broader set of problems, we take Max k-CSP problem as an example. A $\frac{ck}{2^k}$ ($c > 0.44$) approximation bound was obtained for the Max K-CSP \cite{Charikar}. \cite{Charikar} shows this bound by reducing Max k-CSP to Max k-AllEqual, where the approximation bound $\alpha$ for Max k-AllEqual is translated to a $\frac{\alpha}{2}$ bound for Max k-CSP. Thus, we want to show that a $\frac{ck}{2^k}$ ($c > 0.88$) for Max k-AllEqual holds in the robust setting. Let us state definitions of standard and robust Max k-AllEqual problems. 
\begin{definition}
    (Max k-AllEqual) Given a set S of clauses and a set of binary variables $\{x_i \in \{-1, 1\} \}_{i \leq n}$, where each clause C has the form $l_1 \equiv l_2 \equiv \cdots \equiv l_k$, where each literal $l_i$ is either $x_i$ or $\neg x_i$. A clause is satisfied if $l_1 = l_2 = \cdots = l_k = 1$ or $l_1 = l_2 = \cdots = l_k = -1$. The objective is to maximize the number of clauses satisfied. 
\end{definition}
\begin{definition}
    (Robust Max k-AllEqual) Given a weight vector $w \in {\cal W} \subseteq \mathbb{R}_+^{|S|}$, we assign weight $w_C$ to clause $C$. The objective is to maximize the worst-case weighted value of the clauses satisfied. 
\end{definition}
Let us denote the robust Max k-AllEqual problem as (AE) and its optimal value as Val(AE). We consider the SDP relaxation of (AE) as below: 
\begin{equation}
\begin{array}{rl}
\text{Val(AE-R)} \stackrel{\text{def}}{=} \frac{1}{k^2} \max\limits_{\|u_i\|^2 = 1, u_i = -u_{-i}, \forall i \in \{\pm1, \cdots, \pm n \}} \min\limits_{w \in {\cal W}} & \sum\limits_{C \in S} w_C \left\| \sum\limits_{i \in C} u_i \right\|^2
\end{array}
\label{max-allequal-relax}
\tag{AE-R}
\end{equation}
where $u_{-i}$ denotes the negation of $u_i$. To see \eqref{max-allequal-relax} is indeed an outer relaxation of (AE), fix $w \in W$ and let $x^*_1, \ldots, x^*_n$ denote the optimal solution to (AE). We then assign $u_i = e_1$ if $x^*_i = 1$ and assign $u_i = -e_1$ if $x^*_i = -1$. Thus, for each clause $C$, if it is satisfied, $w_C$ is added to the objective value and if it is unsatisfied, $t \cdot w_C$ is added to the objective value where $t \in [0, 1]$. Consequently, Val(AE-R) $\geq$ Val(AE). We now re-state the algorithm in \cite{Charikar} for its robust counterpart.

\begin{algorithm} \label{alg:char-Algorithm}
\caption{Charikar et al. Approximation Algorithm \cite{Charikar} for Robust Max k-AllEqual Problem} 
\begin{algorithmic}[1]
\State \textbf{Step 1:} Solve \eqref{max-allequal-relax}, obtaining optimal solution $\hat{u}_i$ and worst-case weight vector $\hat{w}$. 
\State \textbf{Step 2:} Obtain $\hat{z}_i$ by some polynomial-time rounding algorithm  from $\hat{u}_i$. 
\State \textbf{Step 3:} Independently assign $\hat{x}_i = 1$ with probability $\frac{1 + \sqrt{\frac{2}{k}} \hat{z}_i}{2}$ and $\hat{x}_i = -1$ otherwise. 
\end{algorithmic}
\end{algorithm}
In addition, \cite{Charikar} gives two important theorems: 
\begin{theorem} \cite[Theorem 3.2]{Charikar} \label{thm:char-3.2} There exists a polynomial-time algorithm that given a psd matrix $A$ and a set of unit vectors $\{u_i\}$, assigns $\pm 1$ to variables $z_i$ such that $\sum\limits_{i, j} A_{ij} z_i z_j \geq \frac{2}{\pi} \sum\limits_{i, j} A_{ij} u_i \cdot u_j$.
\end{theorem}
\begin{theorem} \cite[in Proof of Theorem 3.3]{Charikar} \label{thm:char-3.3} For each constraint C, denote $Z_C = \frac{1}{k} \sum\limits_{i \in C}\hat{z}_i$.  Then, $\mathbb{P}(C \text{ is satisfied}) \geq \frac{4\alpha}{e} \cdot \frac{k}{2^k} Z_C^2$, where $\alpha > 0.93945$. 
\end{theorem}

Let $(\hat{u}_i, \hat{w}, \hat{z}_i, \hat{x}_i)$ be obtained from algorithm~4. The following theorem shows that the expected value of the satisfied clauses using algorithm~4 is at least $\frac{0.88k}{2^k}$ times Val(AE). 
\begin{theorem} 
   $\mathbb{E}[\text{value of the satisfied clauses evaluated with weight } w] \geq \frac{0.88k}{2^k}$ Val(AE), $\forall w \in {\cal W}$. 
\end{theorem}
\noindent {\em Proof}. Take arbitrary $w \in {\cal W}$, we have the expectation characterized by
\begin{align*}
    \sum\limits_{C \in S} w_C \mathbb{P}(C \text{ is satisfied}) &\geq \frac{4\alpha}{e} \cdot \frac{k}{2^k} \sum\limits_{C \in S} w_C Z_C^2 \tag{by Theorem \ref{thm:char-3.3}} \\ &= \frac{4\alpha}{e} \cdot \frac{k}{2^k} \cdot \frac{1}{k^2} \sum\limits_{C \in S} w_C \left(\sum\limits_{i \in C}\hat{z}_i\right)^2 \geq \frac{4\alpha}{e} \cdot \frac{k}{2^k} \cdot \frac{2}{\pi} \cdot \frac{1}{k^2} \sum\limits_{C \in S} w_C \left\| \sum\limits_{i \in C} \hat{u}_i \right\|^2 \tag{by Theorem \ref{thm:char-3.2}, assuming nonnegative $w$} \\ &\geq \frac{4\alpha}{e} \cdot \frac{k}{2^k} \cdot \frac{2}{\pi} \cdot \frac{1}{k^2} \sum\limits_{C \in S} \hat{w}_C \left\| \sum\limits_{i \in C} \hat{u}_i \right\|^2 \tag{since $\hat{w}$ is worst} \\ &= \frac{4\alpha}{e} \cdot \frac{k}{2^k} \cdot \frac{2}{\pi} \text{Val(AE-R)} \geq \frac{4\alpha}{e} \cdot \frac{k}{2^k} \cdot \frac{2}{\pi}  \text{Val(AE)} > 0.88 \cdot \frac{k}{2^k} \text{Val(AE)} \tag{by Theorem \ref{thm:char-3.3}}. \qed
\end{align*}

Thus, the $\frac{0.88k}{2^k}$ approximation bound of Max k-AllEqual holds for its robust counterpart, which implies that the $\frac{0.44k}{2^k}$ approximation bound of Max k-CSP also holds for its robust counterpart. Note that this fits into our generalization framework in the robust setting, as the expected value from the algorithm is characterized by the sum of probabilities that can be related to the objective function form of \eqref{max-allequal-relax}, which are independent of the weights. Recent literature has made progress on the optimal inapproximability of satisfiable k-CSPs \cite{Bhangale}. 

\section{Tractability of the Algorithm}
\label{sec:Tractability}
We now discuss the tractability of the G-W algorithm for \eqref{robust-max-cut} and \eqref{dro-max-cut}; specifically, the tractability of \eqref{SDP-MAXCUT} and \eqref{SDP-dro-MAXCUT}. We consider the flattened weight vector $w \in \mathbb{R}^{n^2}$ where $w = \text{vec}(W)$ and decision vector $y \in \mathbb{R}^{n^2}$ where $y = \text{vec}(Y)$. 
\subsection{Robust Maximum Cut with Polyhedral Uncertainty}
Assume the uncertainty set has the formulation ${\cal W} \stackrel{\text{def}}{=} \{w \in \mathbb{R}^{n^2} \,|\, Aw \geq b\}$, for some $A \in \mathbb{R}^{n^2 \times n^2}$ and $b \in \mathbb{R}^{n^2}$. Thus, the inner problem of \eqref{SDP-MAXCUT} is equivalent to 
\begin{equation}
\begin{array}{rl}
\min\limits_{w \in {\cal W}} & \frac{1}{4}(1 - y)^T w
\end{array}
\label{inner-1}
\tag{I}
\end{equation}
which, because of linear programming duality, has a dual formulation
\begin{equation}
\begin{array}{rl}
\max\limits_{p \geq 0, A^Tp = \frac{1}{4}(1 - y)} & b^T p.
\end{array}
\label{inner-dual}
\tag{D-I}
\end{equation}
Incorporating \eqref{inner-dual} into \eqref{SDP-MAXCUT}, we have
\begin{equation}
\begin{array}{rl}
\max\limits_{Y \succeq 0, Y_{ii} = 1, i = 1, \dots, |V|, p \geq 0, A^Tp = \frac{1}{4}(1 - y)} & b^T p.
\end{array}
\label{SDP-MAXCUT-new}
\tag{D-RP-SD}
\end{equation}
\eqref{SDP-MAXCUT-new} is a semidefinite program, which can be solved in polynomial time to some precision $\epsilon$. 

\subsection{Robust Maximum Cut with Ellipsoidal Uncertainty}
Let us now assume an ellipsoidal uncertainty set with the form ${\cal W} \stackrel{\text{def}}{=} \{w \in \mathbb{R}^{n^2} \,|\, (w - w_0)^TQ^{-1}(w - w_0) \leq a\}$, for some $w_0 \in \mathbb{R}^{n^2}$, positive-definite symmetric $Q \in \mathbb{R}^{n^2}$, and positive $a \in \mathbb{R}$. Given the special property of such an ellipsoidal set, \eqref{inner-1} with the new ellipsoidal ${\cal W}$ will admit a closed-form solution $w^* = w_0 - \frac{Q(1-y)}{\sqrt{(1-y)^TQ(1-y)}}$. Thus, \eqref{SDP-MAXCUT} is equivalent to 
\begin{equation}
\begin{array}{rl}
\max\limits_{Y \succeq 0, Y_{ii} = 1, i = 1, \dots, |V|, k} & \frac{1}{4}w_0^T(1-y) - \frac{\sqrt{a}}{4} k \\
\textit{s.t.} & k \leq \lVert Q^{\frac{1}{2}} (1 - y) \rVert_2, -k \leq -\lVert Q^{\frac{1}{2}} (1 - y) \rVert_2.
\end{array}
\label{ellip}
\tag{E-RP-SD}
\end{equation}
Note that \eqref{ellip} is in the form of a semidefinite program, which can also be solved efficiently.

\subsection{Distributionally Robust Maximum Cut with Wasserstein Ambiguity}
Suppose we have an empirical distribution $\mathbb{P}_0$ such that $\mathbb{P}_0(\hat{w}_i) = \frac{1}{n}, \forall i \in [n]$, where $n$ is the number of samples. Let our weight vector $w$ be defined on a measurable space $(\Xi, {\cal F})$, and let ${\cal M}(\Xi, {\cal F})$ be the set of all probability measures in this space, and let $r_0 > 0$ be the radius of our Wasserstein ball. We consider the ambiguity set ${\cal P} \stackrel{\text{def}}{=} \{\mathbb{P} \in {\cal M}(\Xi, {\cal F}) \,|\, {\cal W}(\mathbb{P}, \mathbb{P}_0) \leq r_0\}$, where ${\cal W}(P_1, P_2) \stackrel{\text{def}}{=} \text{inf}_{K \in {\cal S}(P_1, P_2)} \int_{\Xi \times \Xi} d(s_1, s_2) K(ds_1, ds_2)$, where ${\cal S}(P_1, P_2) \stackrel{\text{def}}{=} \{K \in {\cal M}(\Xi \times \Xi, {\cal F} \times {\cal F}) \, | \, K(A \times \Xi) = P_1(A), K(\Xi \times A) = P_2(A), \forall A \in {\cal F}\}$ with a measurable metric $d(\cdot, \cdot)$ on $\Xi$ \cite{GivensShortt1984, Fengqiao}. \cite{Fengqiao} (see also, \cite{GaoKleywegt2022}) provides a reformulation of \eqref{SDP-dro-MAXCUT} into a semi-infinite program by dualizing the inner program: 
\begin{equation}
\begin{array}{rl}
\min\limits_{Y \succeq 0, Y_{ii} = 1, i = 1, \dots, |V|, v_1, \dots, v_m \in \mathbb{R},v_{m + 1} \geq 0} & \frac{1}{n}\sum\limits_{i=1}^{n} v_i + r_0 \cdot v_{m+1} \\
\textit{s.t.} & -\frac{1}{4}(1 - y)^T s - v_i - v_{m + 1} \cdot d(s, \hat{w}_i) \leq 0, s \in \Xi, i \in [n].
\end{array}
\label{SIP-MAXCUT-dro}
\tag{SIP}
\end{equation}
\cite{Fengqiao} also provides a cutting-surface method to solve \eqref{SIP-MAXCUT-dro} to $\epsilon$-optimality in finitely many iterations under the assumption that a separation oracle exists (and possibly solves the problem in polynomial time). The cutting surface approach employs a separation problem to identify a most-violated constraint and add to a finite relaxed version of the semi-infinite \eqref{SIP-MAXCUT-dro}, and continue this procedure until the optimality condition is met.

\section{Conclusion}
\label{sec:ConcludingRemarks}
Among numerous problems in combinatorial optimization, the max cut problem is not only of theoretical interest but has also found many applications in machine learning, physics, and VLSI design, for example, \cite{boykov, frans}. The G-W SDP relaxation and rounding scheme inspired a range of similar techniques applied to other related problems. Furthermore, the approximation bounds are also being improved over time by stronger SDP relaxations and better rounding techniques. For example, following the G-W $0.79607$ approximation bound for Max-DiCut, Feige and Goemans obtained a $0.859$ bound \cite{feige}; Matuura and Matsui obtained a $0.863$ bound \cite{Shiro};  Lewin, Livnat and Zwick obtained a $0.874$ bound \cite{Lewin}, which is the current best approximation bound for Max-DiCut. For another example, the maximum bisection problem also saw a sequence of improving approximation bounds: Frieze and Jerrum, Ye, Halperin and Zwick, Feige and Langberg, Raghavendra and Tan, and Austrin et al. gave approximation bounds of 0.6514, 0.699, 0.7016,  0.7028, 0.85, and 0.8776 respectively \cite{Frieze1997, Ye2001, Halperin, FEIGE20061, Raghavendra, Austrin}. The G-W algorithm also extends to maximum satisfiability, max k-cut, and quantum optimization problems. 

The G-W type rounding algorithm has also been developed for optimization models that are not motivated by combinatorial optimization. For example, worst-case performance bounds were given for quadratically constrained quadratic optimization models under suitable conditions \cite{Nesterov2000, Tseng}. We also refer to \cite{Zhang2000, helmberg2000, Nesterov1998, Ling, Zhang2011} for further such generalizations. Our analysis is also applicable for these generalizations, i.e., the approximation bound for the optimal value in the nominal model will remain valid for their appropriately defined robust and distributionally robust counterparts.

\newpage
\bibliographystyle{informs2014}
\bibliography{annot}

\newpage
\begin{APPENDICES}
  \section{Analysis When the Maximum Cut is Large}
Theorem $3.1.1$ in \cite{Goemans01} discusses the improved approximation ratio when the maximum cut is large. We now re-state their theorem for the robust counterpart \eqref{robust-max-cut}:
\begin{theorem}
    Let $W_{tot} = \sum\limits_{i<j} w_{ij}, \Tilde{A} = \arg\min\limits_{W \in {\cal W}} \frac{1}{W_{tot}} \sum\limits_{i<j} w_{ij} \frac{1 - \hat{u}_i \cdot \hat{u}_j}{2}$, $h(t) = \frac{\arccos(1-2t)}{\pi}$. If $\Tilde{A} \geq \gamma \approx 0.84458$, $\mathbb{E}[\hat{C}(W)] \geq \frac{h(\Tilde{A})}{\Tilde{A}} \text{Val(RP)}, \forall \, W \in {\cal W}$. 
\end{theorem}
\noindent {\em Proof}. Take arbitrary $W \in {\cal W}$, we have
\begin{align*}
    \mathbb{E}[\hat{C}(W)] &\geq \frac{h(\Tilde{A})}{\Tilde{A}} \sum\limits_{i<j} w_{ij} \frac{1 - \hat{u}_i \cdot \hat{u}_j}{2} \tag{follows from G-W's Theorem 3.1.1 \cite{Goemans01}}
    \\ &\geq \frac{h(\Tilde{A})}{\Tilde{A}} \sum\limits_{i<j} \hat{w}_{ij} \frac{1 - \hat{u}_i \cdot \hat{u}_j}{2}
    = \frac{h(\Tilde{A})}{\Tilde{A}} \text{Val(RP-R)}
    \geq \frac{h(\Tilde{A})}{\Tilde{A}} \text{Val(RP)}. \qed
\end{align*}

  \section{Analysis for Negative Edge Weights}
Theorem $3.2.1$ in \cite{Goemans01} discusses the approximation guarantee for arbitrary edge weights. We now re-state their theorem for the robust counterpart \eqref{robust-max-cut}:
\begin{theorem}
    Let $W_{-} = \sum\limits_{i < j: w_{ij} < 0} w_{ij}$. Then, $(\mathbb{E}[\hat{C}(W)] - W_{-}) \geq 0.878(\text{Val(RP)} - W_{-}), \forall \, W \in {\cal W}$.  
\end{theorem}
\noindent {\em Proof}. Take arbitrary $W \in {\cal W}$, we have
\begin{align*}
    (\mathbb{E}[\hat{C}(W)] - W_{-}) &\geq 0.878\left(\sum\limits_{i<j} w_{ij} \frac{1 - \hat{u}_i \cdot \hat{u}_j}{2} - W_{-}\right) \tag{follows from G-W's Theorem 3.2.1 \cite{Goemans01}} \\ &\geq 0.878\left(\sum\limits_{i<j} \hat{w}_{ij} \frac{1 - \hat{u}_i \cdot \hat{u}_j}{2} - W_{-}\right) \\ &= 0.878\left(\text{Val(RP-R)} - W_{-}\right) \geq 0.878\left(\text{Val(RP)} - W_{-}\right). \qed
\end{align*}
\end{APPENDICES}
\end{document}